\input{aipcheck}

\documentclass[
    ,final            % use final for the camera ready runs
  ,sort&compress
  ,numberedheadings % uncomment this option for numbered sections
  ]
  {aipproc}

\layoutstyle{6x9}

\newcommand\aj{AJ} % Astronomical Journal
\newcommand\apj{{ApJ}} % Astrophysical Journal
\newcommand\apjl{{ApJL}} % Astrophysical Journal
\newcommand\apjs{{ApJS}} % Astrophysical Journal, Supplement
\newcommand\aap{{A\&A}} % Astronomy and Astrophysics
\newcommand\pasp{{PASP}} % Astronomy and Astrophysics

\newcommand{\msun}{M$_{\odot}$}

\usepackage{epstopdf}

\begin{document}

\title{Clouds, Brightening and Multiplicity Across the L Dwarf/T Dwarf Transition}


\author{Adam J. Burgasser}{
  address={Massachusetts Institute of Technology}
}

\begin{abstract}
The transition between the two lowest-luminosity spectral classes of brown dwarfs---the L dwarfs and T dwarfs---is traversed by nearly all brown dwarfs as they cool over time.  Yet distinct features of this transition, such as the "J-band bump" and an unusually high rate of multiplicity, remain outstanding problems, although evidence points to condensate cloud evolution as a critical component.  Using a Monte Carlo population simulation that incorporates the empirical spectral properties of unresolved brown dwarfs in magnitude-limited samples, I demonstrate that the J-band bump and enhanced multiplicity naturally emerge from a short timescale of photospheric cloud dissipation.  This timescale may help constrain future evolutionary models exploring the cloud dissipation process.
\end{abstract}

\maketitle

%%%%%%%%%%%%%%%%%%%%%%%%%%%%%%%%%%%%%%%%%%%%
%% MAINMATTER
%%%%%%%%%%%%%%%%%%%%%%%%%%%%%%%%%%%%%%%%%%%%

\section{Context}

The transition between the two coldest classes of brown dwarfs, the L dwarfs and T dwarfs, marks a dramatic change in the chemical abundances, condensate cloud properties and spectral energy distributions of late-type brown dwarfs.  This transition is traversed by nearly all brown dwarfs during their cooling lifetimes.  Yet it remains one of the most poorly understood phases of brown dwarf evolution, both in terms of empirical characterization (relatively few examples are known, most are too faint to be well characterized) and theoretical reproduction of observational data \cite{2002ApJ...571L.151B, 2004AJ....127.3553K, 2008ApJ...678.1372C}.  The L dwarf/T dwarf transition is also distinguished by two notable peculiar features. The ``J-band bump'' is an apparent brightening in the 1.0-1.3 $\mu$m region from late-type L to mid-type T \cite{2002AJ....124.1170D} which is observed as a ``flux reversal'' between the components of L/T binary systems \cite{2006ApJS..166..585B, 2006ApJ...647.1393L, 2008arXiv0803.0544L}.  There is also an excess of binary systems amongst L/T transition objects, up to twice as frequent as warmer L dwarf and cooler T dwarf systems \cite{2006ApJS..166..585B}.  The disappearance of condensate clouds across this transition is likely to be a key factor in  the unusual properties of the L/T transition.  However, given the complexity \cite{2001ApJ...556..872A, 2006A&A...455..325H} and ubiquity of condensate formation in the dynamic, low-temperature atmospheres of both late-type brown dwarfs and exoplanets, the peculiar properties of L/T transition objects can provide important empirical constraints for cloud formation theories.

\section{Simulations}

To understand the origin of the binary excess amongst L/T transition objects, I simulated a volume-complete population of L and T dwarfs, including unresolved binary systems, building off of prior mass function/luminosity function Monte Carlo simulations (\cite{2004ApJS..155..191B}; see Figure~1).  Power law forms of the mass function (dN/dM $\propto$ M$^{-\alpha}$, where $\alpha$ = 0,0.5,1.0,1.5) as well as a lognormal form (parameters from \cite{2002ApJ...567..304C}) were examined.  A constant star formation rate was assumed.  Evolutionary models from both \cite{2001RvMP...73..719B} and \cite{2003A&A...402..701B} were used to convert masses and ages to luminosities (solar metallicity was assumed).  Luminosities were then converted to spectral types using an empirical relation based on single sources \cite{2004AJ....127.3516G} and individual components of resolved binaries \cite{2004A&A...413.1029M, 2005ApJ...634..616L, 2006ApJS..166..585B}.  Simulated binary populations were constructed assuming intrinsic binary fractions ranging from 5--70\%, and both exponential and constant mass ratio distributions were considered \cite{2006ApJS..166..585B}.  Binary spectra were produced by flux-calibrating low-resolution template spectra (obtained with the SpeX spectrograph \cite{2003PASP..115..362R}) according to empirical M$_K$/spectral type relations \cite{2006ApJ...647.1393L,2007ApJ...659..655B}.  The binary spectra, assumed to be unresolved, were classified using calibrated spectral indices.  Space density and binary fraction distributions as functions of spectral type for both volume-limited and magnitude-limited samples (the latter taking into account the overluminosity of unresolved binary systems) were calculated for the full range of parameters examined.

\begin{figure}
  \includegraphics[height=.11\textheight]{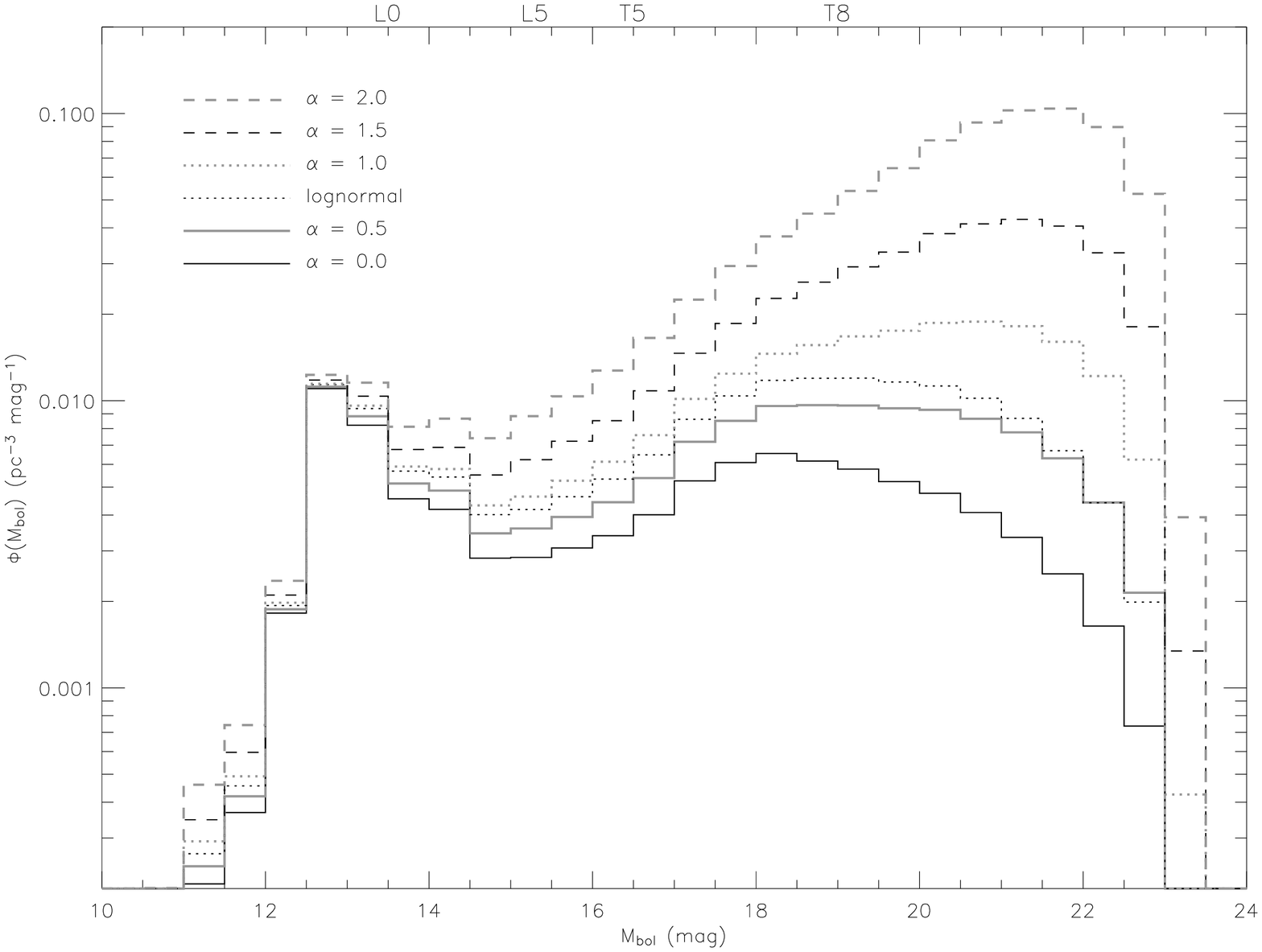}
  \includegraphics[height=.11\textheight]{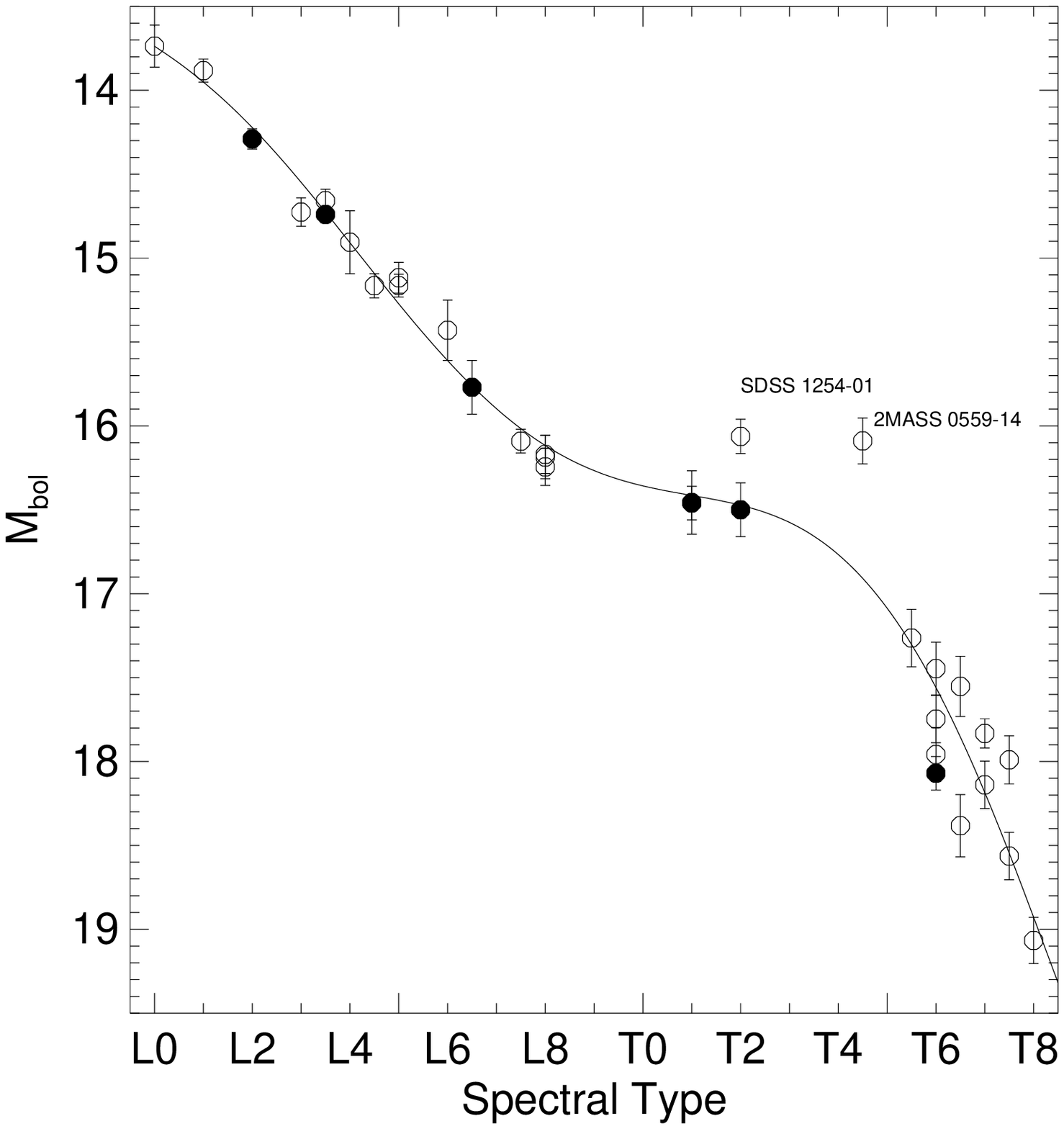}
  \includegraphics[height=.11\textheight]{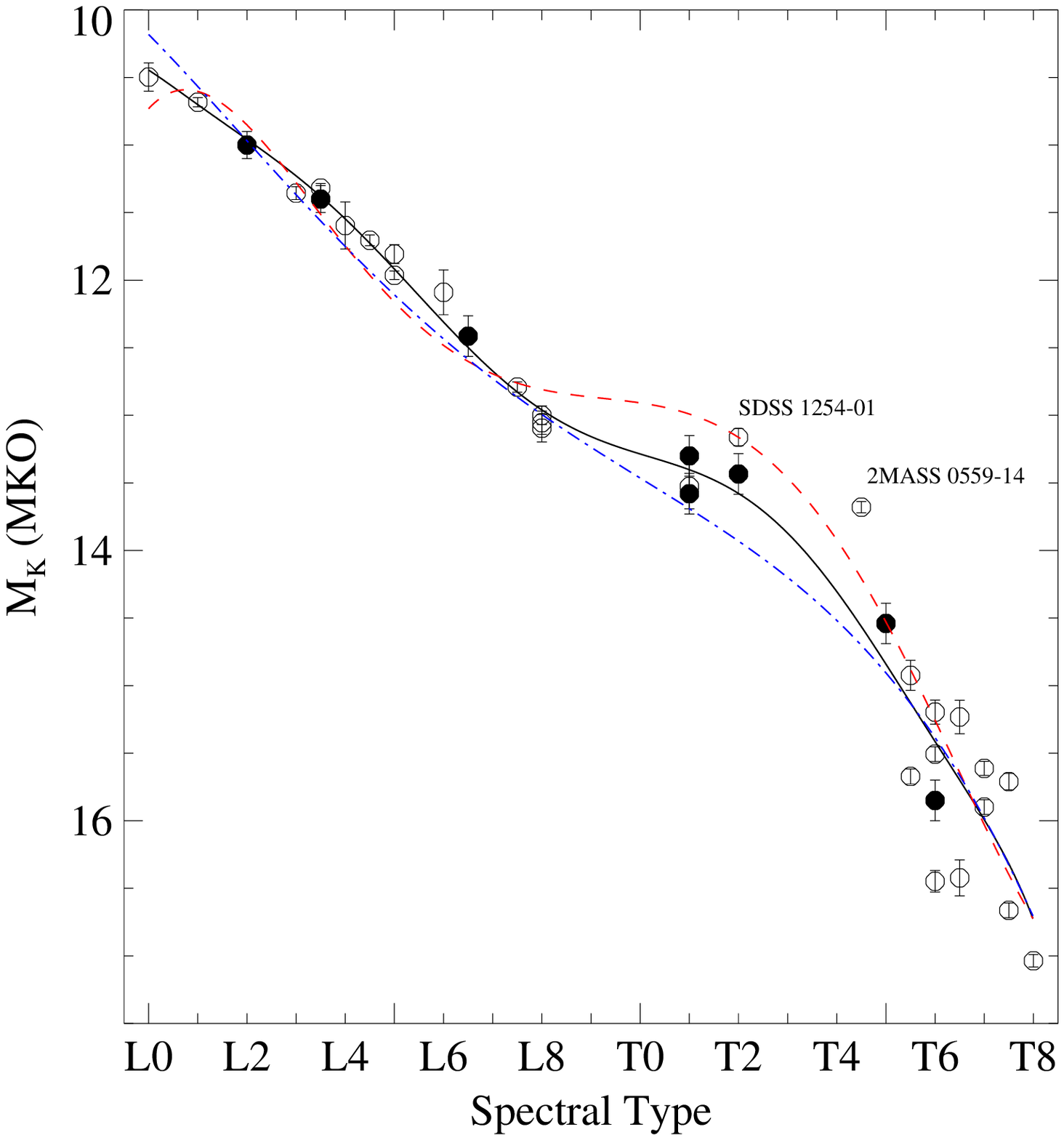}
  \includegraphics[height=.11\textheight]{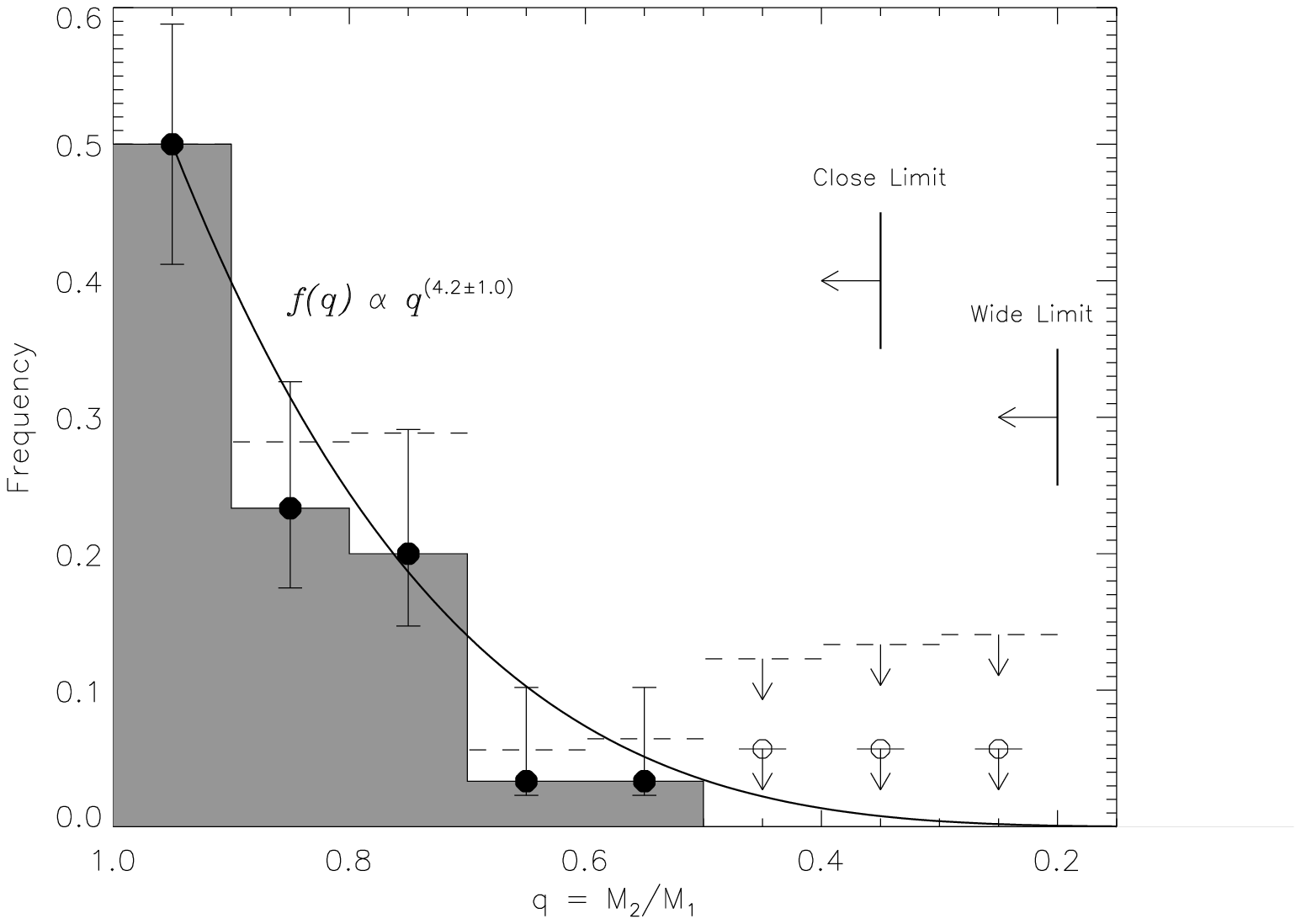}
  \includegraphics[height=.13\textheight]{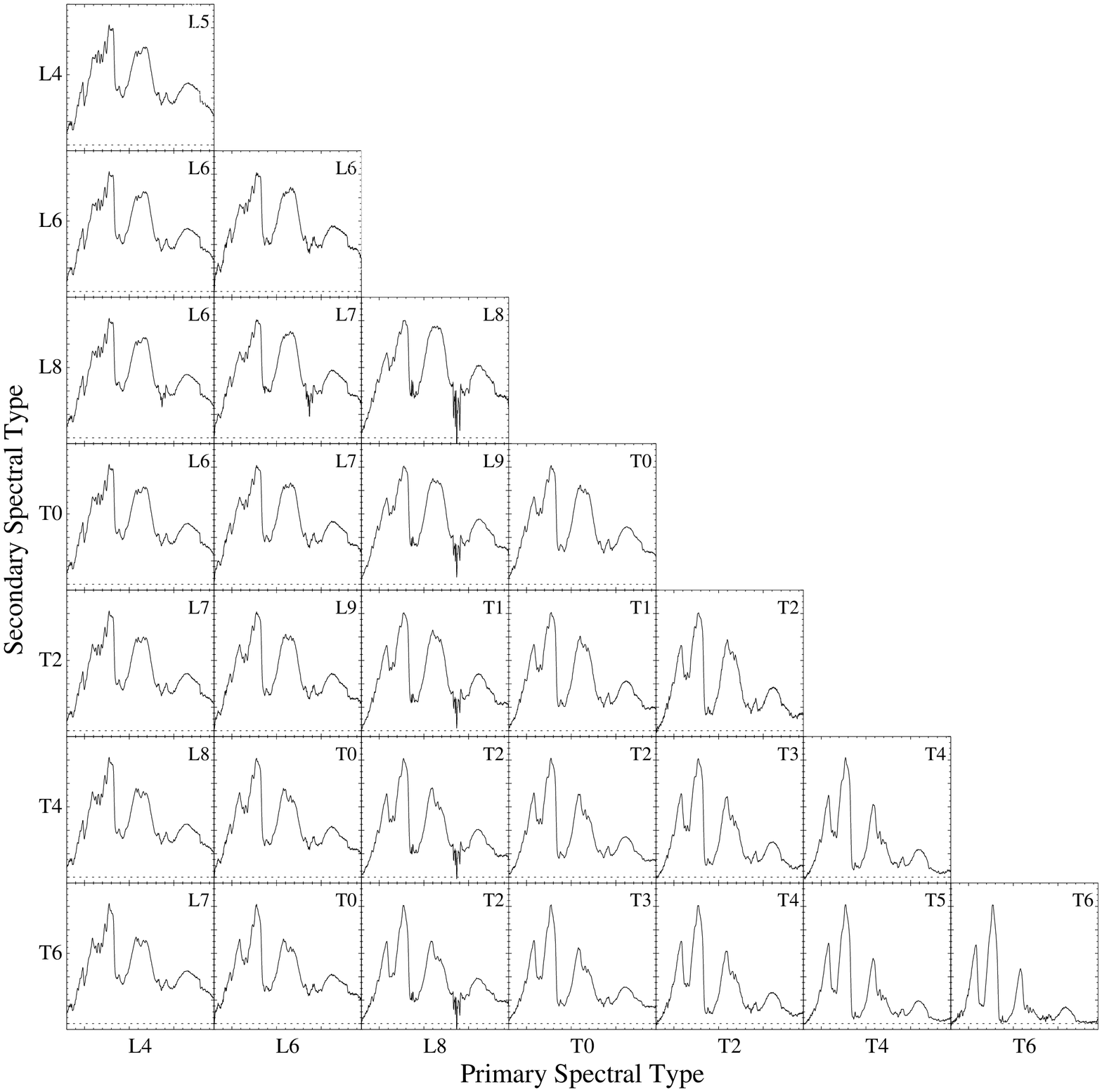}
  \caption{Steps in the L/T binary simulation, from left to right: luminosity function of brown dwarfs based on Monte Carlo mass function simulations, empirical luminosity/spectral type relation, empirical M$_K$/spectral type relation, exponential fit to observed mass ratio distribution of very low mass stars and brown dwarfs, and synthesized combined light (unresolved) binary spectra (from \cite{2004ApJS..155..191B, 2006ApJS..166..585B, 2007ApJ...659..655B}).}
\end{figure}

\section{Results}

The shallow slope of the luminosity/spectral type relation from L7 to T4 (Figure~1) amplifies the dip in the luminosity function seen in prior mass function simulations, predicting $\sim$10$\times$ fewer early-type T dwarfs than other L or T types in a given volume (Figure~2).  The rarity of individual early-type T dwarfs allows them to be outnumbered by hybrid L dwarf + T dwarf pairs, particularly in magnitude-limited samples.

The frequency of binaries as a function of spectral type for both volume-limited and magnitude-limited samples (Figure~2) shows a clear peak at the L/T transition.  For an inherent binary fraction of 11$^{+4}_{-2}$\%, we reproduce the observed (magnitude-limited) resolved binary fraction distribution in detail.  The higher binary fraction of L/T transition objects is attributable to both the paucity of single early-type T dwarfs and the fact that systems comprised of (more common) late-type L dwarf plus T dwarf components resemble early-type T dwarfs.  If the binary fraction across the L/T transition is closer to 66\%, as suggested by  \cite{2006ApJ...647.1393L},  than the intrinsic binary fraction of brown dwarfs may be as high as 40\%, nearly twice current estimates \cite{2006AJ....132..663B,2006ApJS..166..585B}.

The shallow luminosity/spectral type relation inferred from the binary fraction peak implies that brown dwarfs evolve between types L and T---and lose their photospheric condensate clouds in the process---over a relatively short period.  A 0.03 {\msun} brown dwarf makes the jump from L8 to T3 in a mere 100 Myr.  Current equilibrium cloud models predict a much more gradual settling of clouds (e.g., \cite{2002ApJ...568..335M, 2006ApJ...640.1063B}).   Global non-equilibrium effects, reflected in either cloud fragmentation \cite{2002ApJ...571L.151B} or enhanced condensate rain-out  \cite{2004AJ....127.3553K, 2008ApJ...678.1372C} must be an inherent feature of the L/T transition.

\begin{figure}
  \includegraphics[height=.3\textheight]{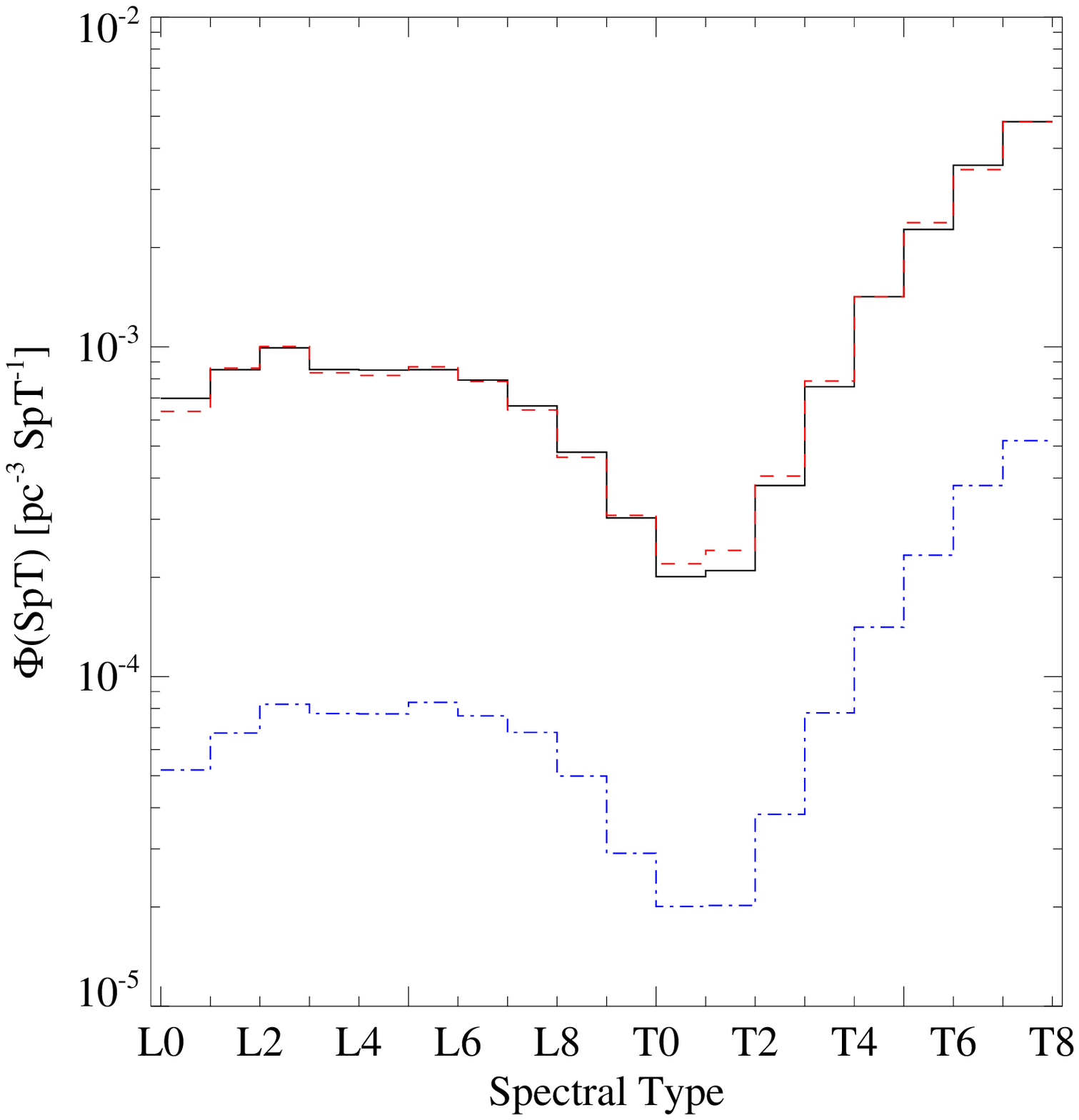}
  \includegraphics[height=.3\textheight]{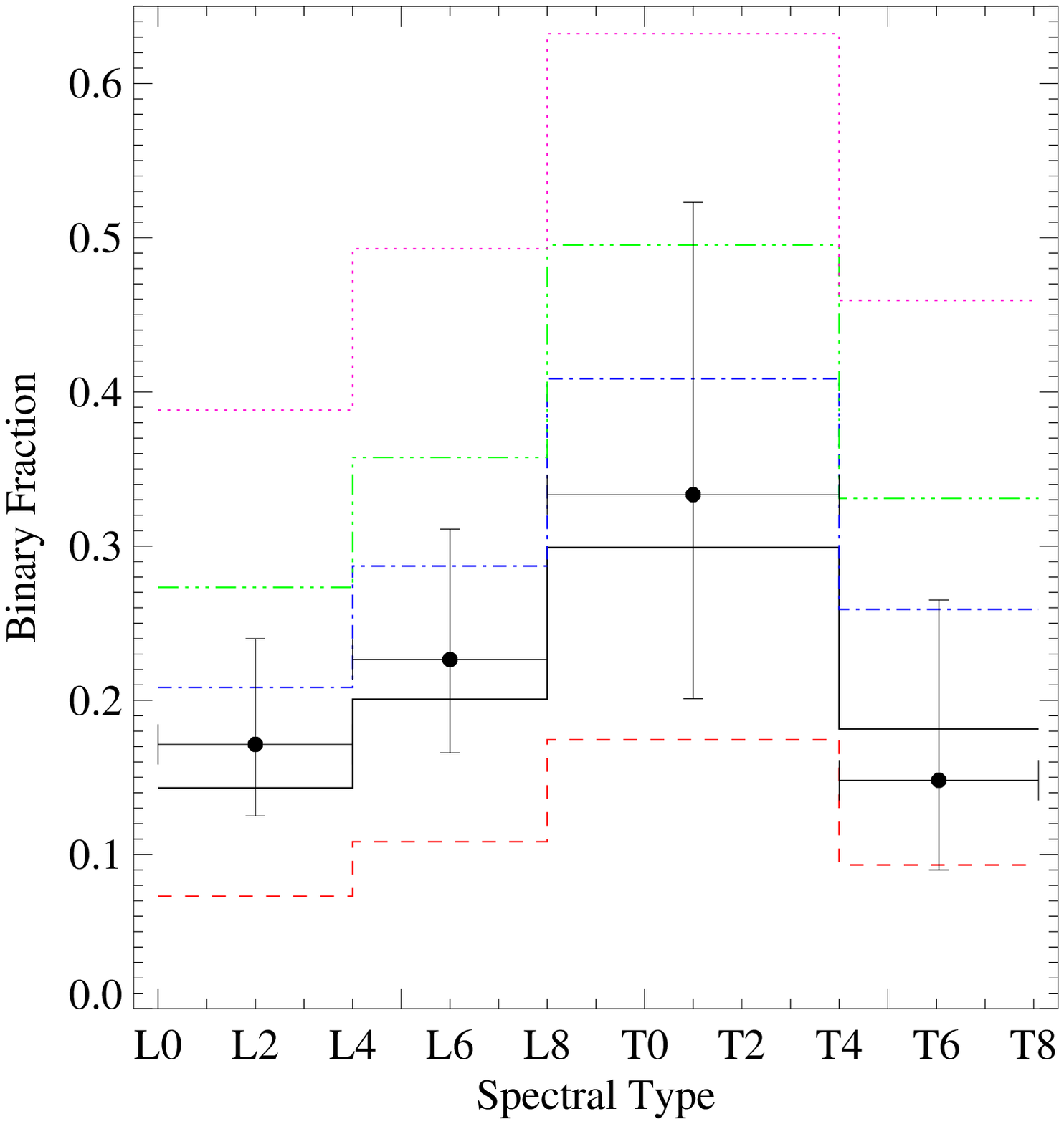}
  \caption{Simulation results. (Left) Volume-limited spectral type distribution of singles (solid black lines), secondaries (dash-dot blue lines) and single + binary systems (dashed red lines).  A deep minimum is found across the L/T transition.  (Right) Observed resolved binary fraction distributions (black points with error bars) compared to predictions for inherent binary fractions of (top to bottom) 5\%, 10\%, 15\%, 20\% and 30\% (from \cite{2007ApJ...659..655B}).}
\end{figure}

\section{New Work}

%The binary excess of L/T transition objects is directly related to the shallow luminosity and absolute magnitude/spectral type relations across the transition, and the corresponding rapidity of condensate cloud loss. 
As binary systems are the best probes of empirical trends across the L/T transition, several groups are attempting to uncover new L/T binaries through resolved imaging studies.  We have recently developed a {\em spectral template matching technique} that identifies and characterizes unresolved binaries from combined-light, low resolution, near-infrared spectroscopy \cite{2007AJ....134.1330B, 2008ApJ...681..579B} (Figure~3).  One of the systems discovered, the M8.5 + T5 dwarf pair 2MASS 0320-0446, has been independently identified as a radial velocity variable \cite{2008ApJ...678L.125B}.  By finding more systems like these, a more robust measure of  the intrinsic brown dwarf binary fraction unaffected by separation limitations is possible, and precise constraints on luminosity and brightness trends across the L/T transition may be made. 

\begin{figure}
  \includegraphics[height=.3\textheight]{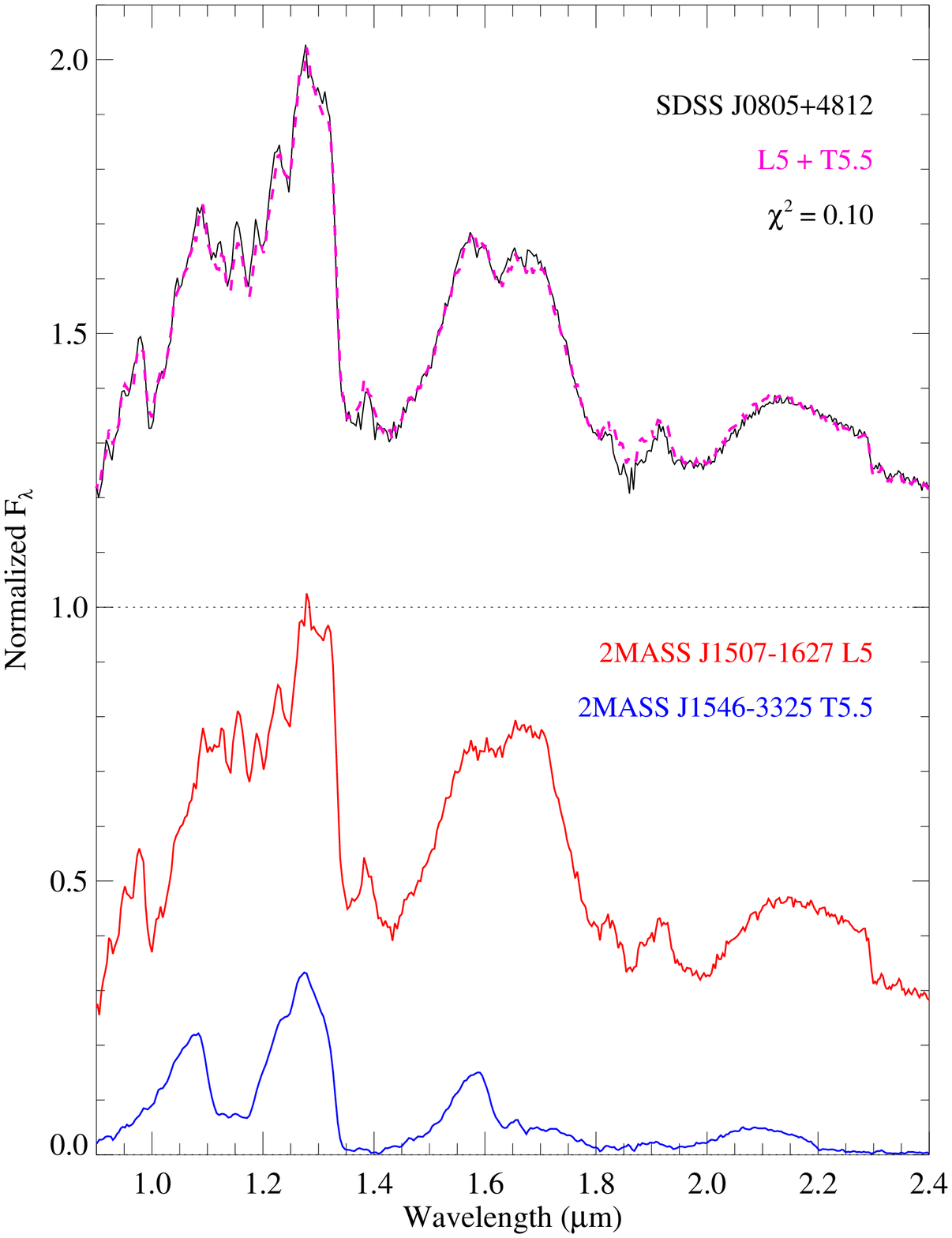}
  \includegraphics[height=.3\textheight]{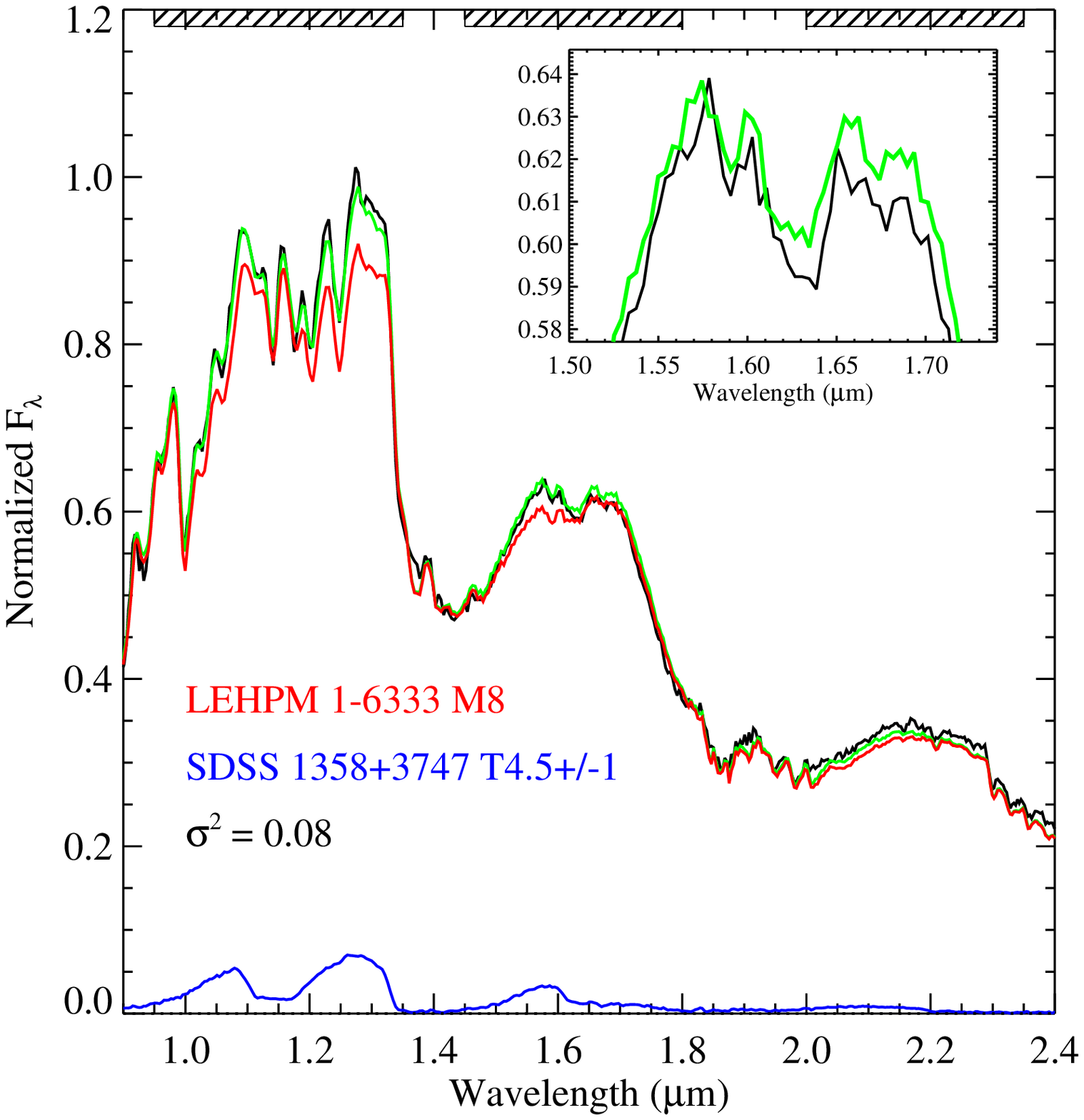}
  \caption{Binaries found by the spectral template matching technique.  (Left) the L4.5 + T5 binary SDSS 0805+4812 \cite{2007AJ....134.1330B}. (Right) the M8.5 + T5 binary 2MASS 0320-0446 \cite{2008ApJ...681..579B}, independently identified as a radial velocity variable \cite{2008ApJ...678L.125B}.}
\end{figure}

%%%%%%%%%%%%%%%%%%%%%%%%%%%%%%%%%%%%%%%%%%%%%%%%
%% BACKMATTER
%%%%%%%%%%%%%%%%%%%%%%%%%%%%%%%%%%%%%%%%%%%%%%%%

%\begin{theacknowledgments}
%HST observations presented here are from program GO-9833.  The author thanks A. Burrows  \& I. Baraffe for electronic versions of theoretical evolutionary models.  This research has made use of Dwarf Archives (http://dwarfarchives.org) maintained by C. Gelino, J.D. Kirkpatrick and A. Burgasser; the VLM Binaries Archive (http://www.vlmbinaries.org) maintained by N. Siegler; and the Spex Prism Libraries (http://www.browndwarfs.org/spexprism) maintained by A. Burgasser.
%\end{theacknowledgments}



\bibliographystyle{aipprocl} % if natbib is missing

%\bibliography{biblibrary}

\IfFileExists{\jobname.bbl}{}
 {\typeout{}
  \typeout{******************************************}
  \typeout{** Please run "bibtex \jobname" to optain}
  \typeout{** the bibliography and then re-run LaTeX}
  \typeout{** twice to fix the references!}
  \typeout{******************************************}
  \typeout{}
 }

\end{document}